\documentclass[aps,prb,reprint,groupedaddress]{revtex4-2}

\usepackage{graphicx}

\begin{document}



\title{On the Applicability Ranges of $T_c$ Formulas for Proximity-Coupled Thin SN and SS Bilayers}


\author{Takayuki Kubo}
\email[]{kubotaka@post.kek.jp}
\affiliation{High Energy Accelerator Research Organization (KEK), Tsukuba, Ibaraki 305-0801, Japan}
\affiliation{The Graduate University for Advanced Studies (Sokendai), Hayama, Kanagawa 240-0193, Japan}



\begin{abstract}
This Brief Note revisits the well-established $T_c$ formulas for proximity-coupled thin superconductor-normal conductor (SN) and superconductor-superconductor (SS) bilayers, highlighting their relationships and clarifying their ranges of applicability. Since these formulas are often misapplied in practical contexts, this note provides guidance for their correct use in experimental and applied settings. For SN bilayers, McMillan's formula is recommended for its broad applicability, with its SS counterpart offering similar reliability.
\end{abstract}

\maketitle



The proximity effect in conventional superconductors has been extensively studied and firmly established through numerous works spanning from the 1960s to the 2000s (e.g., Refs.~\cite{Cooper, McMillan, Khusainov, Golubov_SPIE, Belzig, Fominov, Brammertz}). 
Today, the proximity effect plays a critical role in superconducting device technologies. 
For instance, researchers in the field of transition edge sensors (TES) regularly exploit both superconductor normal conductor (SN) and superconductor superconductor (SS) proximity effects to fine-tune the transition temperature ($T_c$) of bilayer films~\cite{Martinis, Irwin}. 
Moreover, whether intentionally or not, various superconducting devices, such as superconducting cavities for particle accelerators~\cite{snowmass}, are influenced by proximity effects between the device's materials, including the superconducting layers, normal conducting suboxides, impurities, and nonstoichiometric regions on the surface~\cite{Gurevich_Kubo, Kubo_Gurevich}.

This Brief Note focuses on the {\it well-established} $T_c$ formulas for proximity-coupled SN and SS thin bilayers. While these topics are well-studied, with the $T_c$ formula for thin SN bilayers essentially established in McMillan's seminal 1968 work~\cite{McMillan}, revisiting them is crucial to address ongoing confusion in the applied superconductivity communities regarding their improper use. 
Several $T_c$ formulas for thin SN bilayers exist in the literature (e.g., the widely used formula by Martinis et al.~\cite{Martinis}), each with a specific range of applicability. 
However, these formulas are often applied without sufficient consideration of their limits, which can lead to incorrect interpretations of experimental data.

In the following, we briefly summarize the $T_c$ formulas for SN and SS thin bilayers, emphasizing their ranges of applicability and the relationships between them, which have not been well documented and would be valuable for applied communities. 
For the readers' convenience, the derivation of these formulas and the code to calculate $T_c$ are provided in the Supplementary Material.


We consider the geometries shown in Figure 1(a) for the SN bilayer and Figure 1(b) for the SS bilayer. 
The parameters are summarized in Table~\ref{tab:table}. 
As is typical for real device materials, we assume the material is \textit{dirty}, meaning the mean free path is shorter than the intrinsic BCS coherence length. 
In this context, the problem of the proximity effect in SN and SS bilayers is formulated using the quasiclassical Matsubara Green's functions within the Usadel formalism of BCS theory: 
the Usadel equations, the gap equations in regions $i = 1, 2$, and the boundary conditions at the outer surfaces and the interface. 
For simplicity, we consider thin bilayers ($d_{1,2} \lesssim \sqrt{\hbar D_{1,2}/kT_{c}}$). 
Using the well-established techniques of Matsubara summation, we obtain the $T_c$ formula for thin SN and SS bilayers (see the Supplementary Material 1 and Refs.~\cite{Khusainov, Golubov_SPIE, Fominov, Gurevich}).

\begin{figure}[tb]
   \begin{center}
   \includegraphics[height=0.53\linewidth]{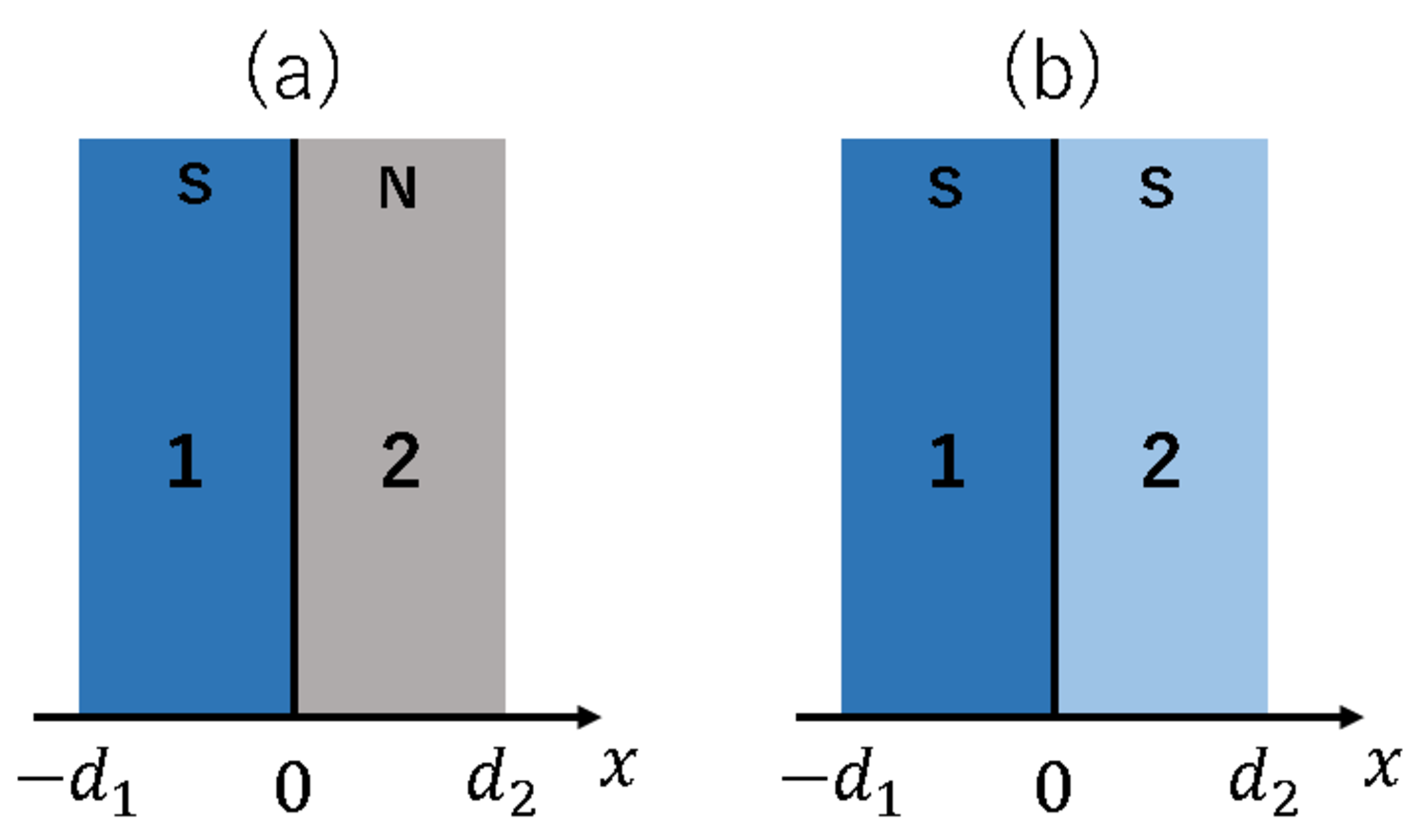}
   \end{center}\vspace{0 cm}
   \caption{
Geometries considered in this article. 
(a) Superconductor-normal conductor thin bilayer. 
(b) Superconductor-superconductor thin bilayer. 
The material parameters are summarized in Table~\ref{tab:table}.
   }\label{f1}
\end{figure}

\begin{table}[t]
\caption{\label{tab:table}
Summary of the parameters. 
}
\begin{tabular}{cc}
\hline
\hline
$T_c$ & Critical temperature of the proximity-coupled bilayer \\ 
$T_{ci}$ & Critical temperature of material $i$ \\ 
$d_i$ & Thickness of material $i$ \\ 
$N_i$ & Normal density of states at the Fermi energy in material $i$ \\ 
$\Theta_{i}$ & Debye temperature of material $i$ \\ 
$R_B$ & Interface resistance between materials 1 and 2 \\ 
\hline
\hline
\end{tabular}
\end{table}

The {\it general} $T_c$ formula for a thin \textit{SN bilayer} is well-established and given by~\cite{Khusainov, Fominov, Gurevich} (with the derivation provided in the Supplementary Material 1):
\begin{eqnarray}
\ln \frac{T_c}{T_{c1}} = \frac{\alpha}{1+\alpha} \left\{ \ln \left( 1 + \frac{1+\alpha}{ (\Theta_{1}/T_{c1}) \alpha \beta} \right) - u(T_c) \right\}, \label{SN_general}
\end{eqnarray}
where the function $u(T_c)$ is defined using the digamma function $\psi$ as follows:
\begin{eqnarray}
u(T_c) = \psi \left( \frac{1}{2} + \frac{1+\alpha}{2\pi (T_c/T_{c1}) \alpha \beta} \right) - \psi \left( \frac{1}{2} \right).
\end{eqnarray}
The material parameters summarized in Table~\ref{tab:table} are aggregated into the following dimensionless parameters:
\begin{eqnarray}
&&\alpha = \frac{N_2 d_2}{N_1 d_1}, \\
&&\beta = 4 \frac{e^2}{\hbar} R_B k T_{c1} N_1 d_1.
\end{eqnarray}
Note that our parameterization is similar to that of Gurevich~\cite{Gurevich}, but our $\beta$ contains the additional factor $kT_{c1}/\hbar$ to ensure $\beta$ is dimensionless.
The transition temperature of thin SN bilayers can be obtained by numerically solving Eq.~(\ref{SN_general}) with respect to $T_c/T_{c1}$. 
This is a straightforward problem.

It is important to note that in deriving the formula [Eq.~(\ref{SN_general})], the Kuprianov-Lukichev boundary conditions~\cite{BC} were applied, as is standard in the referenced works~\cite{Khusainov, Golubov_SPIE, Belzig, Fominov, Brammertz, Gurevich_Kubo, Kubo_Gurevich, Martinis, Irwin, Gurevich} (see also the Supplementary Material 1). 
These boundary conditions are derived from the more general formulations~\cite{Nazarov, Linder} for interfaces with low transmission coefficients ($t < 1$), which in our parametrization corresponds to $d_1/\xi_0 < \beta < \infty$. 
Here, $\xi_0 = \hbar v_f / \pi \Delta_0$ is the intrinsic BCS coherence length in the superconductor layer, and is typically large in conventional metallic superconductors used in devices (e.g., $\xi_0^{\rm (Al)} > 1\,{\rm \mu m}$, $\xi_0^{\rm (Ti)} \simeq 300\,{\rm nm}$, $\xi_0^{\rm (Nb)} \simeq 40\,{\rm nm}$). 
As a result, the applicability range of Eq.~(\ref{SN_general}) spans a wide range of $\beta$, from $\beta \ll 1$ to $\beta \gg 1$. 
In the following, we assume this condition is always satisfied, and the general formula [Eq.~(\ref{SN_general})] remains valid throughout.

\begin{figure}[tb]
   \begin{center}
\includegraphics[width=0.97\linewidth]{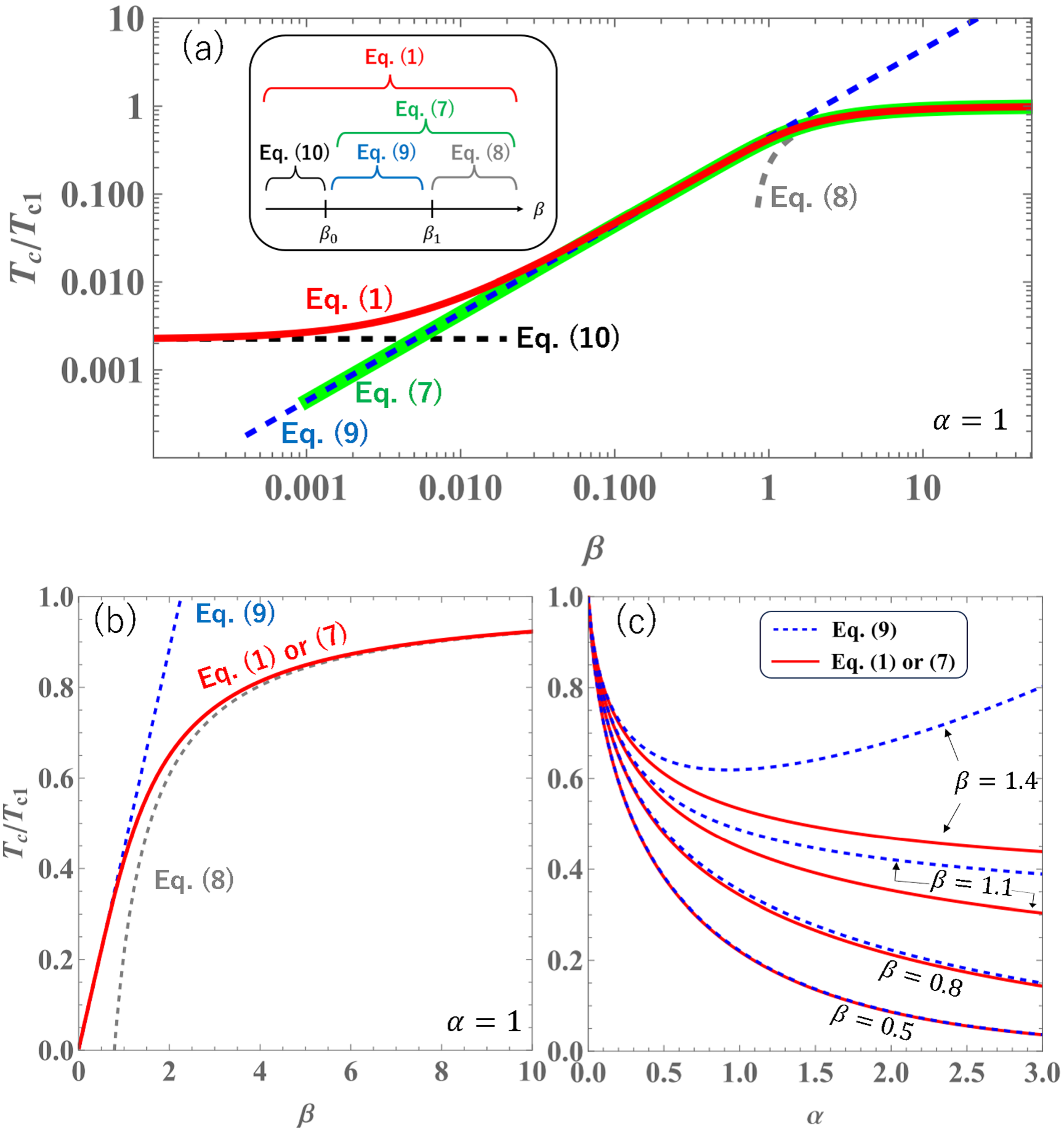}
   \end{center}\vspace{0 cm}
   \caption{
(a) Transition temperature of a thin superconducting-normal (SN) bilayer calculated from different $T_c$ formulas. 
The inset shows a schematic view of the applicable range of each formula. 
(b) Transition temperature of a thin superconducting-normal (SN) bilayer as a function of the dimensionless interface barrier parameter, $\beta$, with a fixed dimensionless thickness parameter, $\alpha = 1$. 
(c) Transition temperature of a thin SN bilayer as a function of the dimensionless thickness parameter, $\alpha$, for various values of the interface barrier parameter, $\beta$.
   }\label{f2}
\end{figure}

Based on the magnitude of the dimensionless interface resistance $\beta$, we can categorize the SN proximity effect into three different regimes: small $\beta$ ($\beta \ll \beta_0$), large $\beta$ ($\beta \gg \beta_1$), and moderate $\beta$ ($\beta_0 \ll \beta \ll \beta_1$). 
Here, $\beta_0$ and $\beta_1$ are given by:
\begin{eqnarray}
&&\beta_0 = \left( 1+\frac{1}{\alpha} \right) \frac{T_{c1}}{\Theta_1}, \\
&&\beta_1 = \frac{\pi}{4} + \frac{1}{2\pi} + \frac{1}{2\pi\alpha} .
\end{eqnarray}
Note that the Debye temperature is typically $\Theta_1 \sim 100\,{\rm K} \gg T_{c1}$. 
For instance, assuming $\alpha \sim 1$, we find $\beta_0 \sim 10^{-2}$ and $\beta_1 \sim 1$, resulting in $\beta_0 \ll \beta_1$.

In typical situations, the interface resistance is sufficiently large such that $\beta \gg \beta_0$. In this regime, we can neglect the logarithmic term in the parentheses, and Eq.~(\ref{SN_general}) simplifies to
\begin{eqnarray}
\ln \frac{T_c}{T_{c1}} = -\frac{\alpha}{1+\alpha} u(T_c) \hspace{0.5cm} (\beta_0 \ll \beta), 
\label{SN_McMillan}
\end{eqnarray}
which is the formula derived by McMillan in 1968 [see Eq.~(39) of Ref.~\cite{McMillan}]. 
In this case, $T_c$ depends only on $\alpha$ and $\beta$; the parameter $\Theta_1/T_{c1}$ is not required. 
Further assuming $\beta \gg (1 + 1/\alpha) T_{c1}/2\pi T_c = \beta_1$, the first term in $u(T_c)$ can be expanded around $\psi(1/2)$, giving the formula:
\begin{eqnarray}
\frac{T_c}{T_{c1}} = 1 - \frac{\pi}{4\beta} \hspace{0.5cm} (\beta_1 \ll \beta), 
\label{SN_large_beta}
\end{eqnarray}
On the other hand, when $\beta \ll \beta_1$, using the asymptotic form of the digamma function, we have $u \simeq \ln [(2e^{\gamma_E}/\pi)(1+\alpha) / \{ (T_c/T_{c1})\alpha\beta\} ]$, where $\gamma_E = 0.577$ and $2e^{\gamma_E}/\pi = 1.13$. 
Then, Eq.~(\ref{SN_McMillan}) reduces to
\begin{eqnarray}
\frac{T_c}{T_{c1}} = \left( \frac{\beta}{1.13 (1 + 1/\alpha)} \right)^{\alpha} \hspace{0.5cm} (\beta_0 \ll \beta \ll \beta_1), 
\label{SN_moderate_beta}
\end{eqnarray}
which corresponds to the formula shown by Martinis et al. [see Eq.~(9) of Ref.~\cite{Martinis}].

In the opposite limit to Eqs.~(\ref{SN_McMillan})-(\ref{SN_moderate_beta}), we consider the case where $\beta \to 0$. In this limit, Eq.~(\ref{SN_general}) simplifies to:
\begin{eqnarray}
\frac{T_c}{T_{c1}} = e^{-\frac{1}{g_1} \frac{N_2 d_2}{N_1 d_1}} \hspace{0.5cm} (\beta \to 0),
\label{SN_Cooper}
\end{eqnarray}
Here, $g_1 = 1/\ln(1.13\Theta_1/T_{c1})$ is the BCS coupling constant in region 1 (the superconductor side). 
Eq.~(\ref{SN_Cooper}), obtained by Cooper in 1961~\cite{Cooper}, is known as the Cooper limit.
It is worth noting that Eq.~(\ref{SN_McMillan}), as well as Eqs.~(\ref{SN_large_beta}) and (\ref{SN_moderate_beta}), do not reproduce the Cooper limit even at vanishing interface resistance ($\beta \to 0$). 
The reason is clear~\cite{Fominov}: 
Eqs.~(\ref{SN_McMillan})-(\ref{SN_moderate_beta}) are valid for moderate to large interface resistance. 
To reproduce Cooper's result, the logarithmic term in the parentheses in Eq.~(\ref{SN_general}) is necessary.

Figure~\ref{f2}(a) shows the $T_c$ of a thin SN bilayer as a function of the dimensionless interface barrier $\beta$, comparing different $T_c$ formulas. 
The inset illustrates the applicable range of each formula. 
Unlike Eq.~(\ref{SN_general}) and McMillan's formula [Eq.~(\ref{SN_McMillan})], the other formulas [Eqs.~(\ref{SN_large_beta})-(\ref{SN_Cooper})] cover narrower ranges of the parameter space.

Figure~\ref{f2}(b) presents linear plots of $T_c$ as a function of $\beta$. 
Eq.~(\ref{SN_general}) and McMillan's formula [Eq.~(\ref{SN_McMillan})] show excellent agreement across the entire range. 
In contrast, Eqs.~(\ref{SN_large_beta}) and (\ref{SN_moderate_beta}) produce {\it unphysical} results (i.e., $T_c < 0$ and $T_c > T_{c1}$, respectively) when applied outside their valid ranges.

Figure~\ref{f2}(c) shows linear plots of $T_c$ as functions of the dimensionless N layer thickness $\alpha$. 
The widely used formula, Eq.~(\ref{SN_moderate_beta}), provides good approximations of $T_c$ for $\beta \lesssim \beta_1 \sim 1$, as before. 
However, outside of its applicable range ($\beta \gtrsim \beta_1 \sim 1$), it yields {\it unphysical} results: $T_c$ {\it increases} with $\alpha$.

Typical $T_c$ values for thin film devices fall within the range $0.1 < T_c/T_{c1} < 1$ for $\alpha \sim 1$, suggesting $\beta \gtrsim 0.1$. 
These $\beta$ values are well-covered by both Eq.~(\ref{SN_general}) and McMillan's formula [Eq.~(\ref{SN_McMillan})]. 
While Eqs.~(\ref{SN_large_beta}) and (\ref{SN_moderate_beta}) can also be applied, it is essential to consider their more limited ranges of applicability. 
Applying these formulas outside their valid ranges results in {\it unphysical results}. 
For greater reliability, McMillan's formula [Eq.~(\ref{SN_McMillan})] is recommended, and the corresponding code is provided in the Supplementary Material 2.

\begin{figure}[tb]
   \begin{center}
   \includegraphics[width=0.96\linewidth]{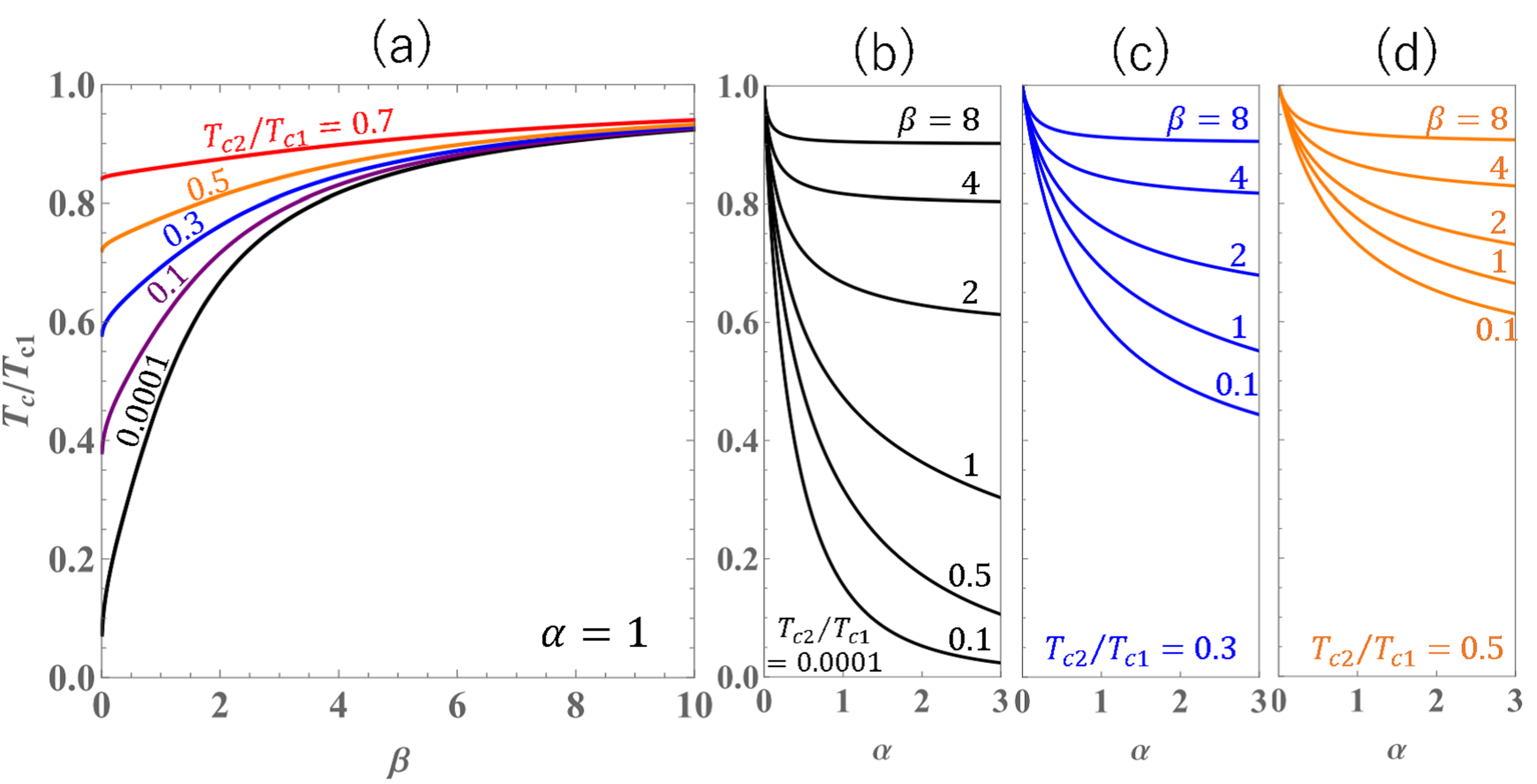}
   \end{center}\vspace{0 cm}
   \caption{
(a) Transition temperature $T_c$ of thin SS bilayers as a function of $\beta$ for various $T_{c2}/T_{c1}$ ratios, representing different material combinations. 
(b, c, d) $T_c$ of thin SS bilayers as a function of the dimensionless thickness $\alpha$ for different values of $\beta$, with material combinations corresponding to $T_{c2}/T_{c1} = 0.0001$, $0.3$, and $0.5$, respectively.
   }\label{f3}
\end{figure}


For completeness, we consider a thin {\it SS bilayer} [see Figure~\ref{f1}(b)]. 
Its $T_c$ formula is given by Eq.~(S18) of the the Supplementary Material 1, 
which is the counterpart to the SN $T_c$ formula, Eq.~(\ref{SN_general}).
This formula, Eq.~(S18), simplifies significantly when the dimensionless interface resistance $\beta$ satisfies $\beta \gg \max \{ \beta_0, \beta_0(\Theta_1/\Theta_2) \}$, resulting in
\begin{eqnarray}
\Bigl( \ln \frac{T_c}{T_{c1}} \Bigr)^2 + \bigl\{ u(T_c) -\ln r \bigr\} \ln\frac{T_c}{T_{c1}} - \frac{\alpha \ln r}{1+\alpha} u(T_c)=0 , \label{SS_McMillan}
\end{eqnarray}
where $r = T_{c2}/T_{c1}$. 
Eq.~(\ref{SS_McMillan}) is the SS counterpart to McMillan's SN $T_c$ formula [Eq.~(\ref{SN_McMillan})], and it provides reliable and practical results over a wide range of parameter space. 
Examples of the calculated results are presented in Figure~\ref{f3}. 
The corresponding code is provided in the Supplementary Material 2.

In summary, this Brief Note revisited the well-established $T_c$ formulas for proximity-coupled SN and SS thin bilayers, emphasizing their applicability and interrelationships. 
The five $T_c$ formulas for a thin SN bilayer are visually summarized in Fig.~\ref{f2}(a), offering a convenient reference for readers seeking to quickly understand the applicability range of each formula. 
For thin SN bilayers, McMillan's formula [Eq.~(\ref{SN_McMillan})] is recommended due to its broad applicability. 
For thin SS bilayers, Eq.~(\ref{SS_McMillan}), the SS counterpart to McMillan's formula, is similarly recommended for its wide applicability. 
The derivation of the $T_c$ formulas and the corresponding code are provided in the Supplementary Material for convenience.


\begin{acknowledgments}
I am deeply grateful to everyone who generously supported my extended paternity leave, which lasted three years. Your support allowed me to spend invaluable time with my family~\cite{ikuji}. 
This work originated from a project on the proximity effect in superconducting cavities for particle accelerators during my leave, a part of which evolved into this brief note following insightful discussions with superconducting device researchers.  
The primary results of the project during my leave will be presented elsewhere.
This work was supported by JSPS KAKENHI Grants No. JP17KK0100 and Toray Science Foundation Grants No. 19-6004.
\end{acknowledgments}


\begin{thebibliography}{99}


\bibitem{Cooper}
L. N. Cooper, 
Superconductivity in the Neighborhood of Metallic Contacts, 
Phys. Rev. Lett. {\bf 6}, 689 (1961). 

\bibitem{McMillan}
W. L. McMillan, 
Tunneling Model of the Superconducting Proximity Effect,
Phys. Rev. {\bf 175}, 537 (1968). 

\bibitem{Khusainov}
M. G. Khusainov, 
Proximity effect with arbitrary transmission of the NS boundary, 
JETP Lett. {\bf 53}, 579 (1991). 

\bibitem{Golubov_SPIE}
A. Golubov, 
Proximity effect in dirty N/S multilayers, 
in {\it Superconducting Superlattices and Multilayers}, edited by I. Bozovic, SPIE Proceedings, Vol. 2157 (SPIE, Bellingham, 1994), p353. 

\bibitem{Belzig}
W. Belzig, C. Bruder, and G. Schon, 
Local density of states in a dirty normal metal connected to a superconductor, 
Phys. Rev. B {\bf 54}, 9443 (1996). 

\bibitem{Fominov}
Y. V. Fominov and M. V. Feigel'man, 
Superconductive properties of thin dirty superconductor-normal-metal bilayers, 
Phys. Rev. B {\bf 63}, 094518 (2001). 

\bibitem{Brammertz}
G. Brammertz, A. Poelaert, A. A. Golubov, P. Verhoeve, A. Peacock, and H. Rogalla, 
Generalized proximity effect model in superconducting bi- and trilayer films, 
J. Appl. Phys. {\bf 90}, 355 (2001). 


\bibitem{Martinis}
J. M. Martinis, G. C. Hilton, K. D. Irwin, and D. A. Wollman, 
Calculation of $T_c$ in a normal-superconductor bilayer using the microscopic-based Usadel theory, 
Nuclear Instruments and Methods in Physics Research A {\bf 444}, 23 (2000).

\bibitem{Irwin}
K. D. Irwin and G. C. Hilton, 
Transition-Edge Sensors, 
in {\it Cryogenic Particle Detection}, vol 99 (Springer, Berlin, Heidelberg, 2005), p. 63. 

\bibitem{snowmass}
A. Gurevich, T. Kubo, and J. A. Sauls, 
in {\it Proceedings of the 2021 US Community Study on the Future of Particle Physics (Snowmass 2021), WA, USA} (APS Division of Particles and Fields, MD, USA, 2022),
https://www.slac.stanford.edu/econf/C210711/

\bibitem{Gurevich_Kubo}
A. Gurevich and T. Kubo, 
Surface impedance and optimum surface resistance of a superconductor with an imperfect surface, 
Phys. Rev. B {\bf 96}, 184515 (2017). 

\bibitem{Kubo_Gurevich}
T. Kubo and A. Gurevich, 
Field-dependent nonlinear surface resistance and its optimization by surface nanostructuring in superconductors, 
Phys. Rev. B {\bf 100}, 064522 (2019). 


\bibitem{Gurevich}
A. Gurevich, 
Tuning vortex fluctuations and the resistive transition in superconducting films with a thin overlayer, 
Phys. Rev. B {\bf 98}, 024506 (2018). 


\bibitem{BC}
M. Yu. Kuprianov and V. F. Lukichev, 
Sov. Phys. JETP {\bf 67}, 1163 (1988). 

\bibitem{Nazarov}
Yu. V. Nazarov, 
Novel circuit theory of Andreev reflection, 
Superlattices Microstruct. 25, 1221 (1999).

\bibitem{Linder}
M. Eschrig, A. Cottet, W. Belzig, and J. Linder, 
General boundary conditions for quasiclassical theory of superconductivity in the diffusive limit: application to strongly spin-polarized systems,
New J. Phys. {\bf 17}, 083037 (2015).

\bibitem{ikuji}
T. Kubo, 
An Encouraging of Paternity Leave: A Physicist Who Has Become a Stay-at-Home Dad in New York, 
KASOKUKI, {\bf 20}, 50 (2023) [Journal of the Particle Accelerator Society of Japan {\bf 20}, 50 (2023)].



\end{thebibliography}
\end{document}